\documentclass[prd,twocolumn,showpacs,preprintnumbers,superscriptaddress,floatfix,nofootinbib]{revtex4-1}
\usepackage{amssymb}
\usepackage{amsmath,txfonts}
\usepackage{graphicx}
\usepackage{dcolumn}
\usepackage{color}
\usepackage{bm}
\usepackage[subfigure]{graphfig}
\usepackage{makecell}
\usepackage[colorlinks,
            citecolor=blue,
            anchorcolor=green,
            menucolor=orange,
            linkcolor=red,
            filecolor=red,
            runcolor=pink,
            urlcolor=blue,
            frenchlinks=red]{hyperref}

\begin{document}

\title{\boldmath Analysis of recent CLAS data on $f_{1}(1285)$
photoproduction}
\author{Xiao-Yun Wang}
\thanks{xywang@lut.cn}
\affiliation{Department of physics, Lanzhou University of Technology,
Lanzhou 730050, China}
\author{Jun He}
\thanks{Corresponding author : jun.he.1979@icloud.com}
\affiliation{Department of  Physics and Institute of Theoretical Physics, Nanjing Normal University,
Nanjing, Jiangsu 210097, China}
\affiliation{Theoretical Physics Division, Institute of Modern Physics,
Chinese Academy of Sciences, Lanzhou 730000, China}
\begin{abstract}
Based on the experimental data released recently by the CLAS Collaboration,
the $f_{1}(1285)$ photoproduction off a proton target is investigated in an
effective Lagrangian approach. In our model, $s$-channel, $u$-channel, and $%
t $-channel Born terms are included to calculate the differential cross
sections, which are compared with the recent CLAS experiment. An interpolating
Reggeized treatment is applied to the $t$ channel, and it is found that the $%
t$-channel contribution is dominant in the $f_1(1285)$ photoproduction and
the $u$-channel contribution is responsible for  the enhancement at backward
angles. The present calculation does not take into account any explicit $s$%
-channel nucleon-resonance contribution; nevertheless, our model provides a
good description of the main features of the data. It suggests small
couplings of the $f_1(1285)$ and the nucleon resonances in this energy
region.
\end{abstract}

\maketitle

\section{Introduction}

Thanks to  high-quality photoproduction data obtained at facilities with
electromagnetic probes, great progress has been achieved in the study of the
hadron spectroscopy, especially of the nucleon resonance, during the last
few decades. Meson photoproduction off a baryon provides one of the most
direct routes to extract information regarding the hadronic structure. At present,
the nature of nucleon resonances below 2 GeV has been widely investigated in
both experiment and theory. However, studies of the properties of nucleon
resonances above 2 GeV are somewhat scarce, and there exist many problems to explain their internal structure~\cite{He:2013ksa,He:2014gga,He:2015yva,He:2017aps}.

The $f_{1}(1285)$ photoprodcution attracts special attention due to its
large threshold of production energy, which provides an opportunity to study
the properties of nucleon resonance above 2 GeV. Furthermore, the nature of
the $f_{1}(1285)$ is also an interesting topic in the hadronic spectroscope
and has been studied for many years. In the Review of Particle Physics (PDG)
the $f_{1}(1285)$ is an axial-vector state with quantum number $%
I^{G}(J^{PC})=0^{+}(1^{++})~$\cite{Olive:2016xmw}. In Refs. \cite{Roca:2005nm,Lutz:2003fm}, the $%
f_{1}(1285)$ was suggested to be a dynamically generated state produced from
the $K\bar{K}^{\ast }$ interaction. In addition, the $f_{1}(1285)$ appears
as a bound state in the dynamical picture within the frame of chiral unitary
approach in Refs. \cite{Geng:2015yta,Zhou:2014ila}. Recently, a calculation in
the one-boson-exchange model suggested that the $f_{1}(1285)$ is the strange
partner of the $X(3872)$ in the hadronic molecular state picture~\cite%
{Lu:2016nlp}. Thus, investigation of the $f_{1}(1285)$ photoproduction may
provide useful information for better understanding the nature of the $%
f_{1}(1285)$. Besides its decay pattern extracted in its decay process, we
may obtain the information about the decay of a nucleon resonance to the $%
f_1(1285)$ and a nucleon and its radiative decay to vector meson.

In the past, due to lack of experimental data, studies of the $f_{1}(1285)$
photoproduction off a proton are scarce, except for some theoretical predictions where only the $t$ channel is considered~\cite{Kochelev:2009xz,Domokos:2009cq}, before the
recently released CLAS data~\cite{Dickson:2016gwc}. Though these theoretical models
provide a basic frame for the $f_1(1285)$ photoproduction and are important
to push forward the experimental studies, they fail to fit the new CLAS
data~\cite{Dickson:2016gwc}. In the new data released by the CLAS Collaboration, the
differential cross sections were measured from threshold up to a
center-of-mass energy of 2.8 GeV in a wide range of production angle. The
cross section falls off in the forward-most angle bins, which is not typical
in meson photoproduction~\cite{Dickson:2016gwc}, which suggests a $t$-channel
contribution with an interpolating Reggeized treatment~\cite{Nam:2010au}.
Moreover, new CLAS data also provide results at backward angles where an
enhancement can be found, which indicates the possible $u$-channel
contribution~\cite{He:2013ksa,He:2014gga}. In the experimental paper~\cite%
{Dickson:2016gwc}, a helicity system fit to the $\eta\pi^+\pi^-$ Dalitz distribution
was done,  and it  found that an $s$-channel nucleon resonance with spin parity $J^P=3/2^+$
instead of the $t$ channel is dominant in the $f_1(1285)$ photoproduction
mechanism, but there does not exist such a candidate in the PDG.

The CLAS data provide an opportunity to understand the reaction mechanism of the
$f_1(1285)$ photoproduction. It is interesting to make a more explicit
analysis of the new CLAS data. In this work, we will analyze the $%
f_1(1285)$ photoproduction based on the CLAS data in an effective Lagrangian
approach. Besides the $t$ channel, which has been considered in the predictions
of Ref.~\cite{Kochelev:2009xz}, the $u$ channel and $s$ channel will also be included
in our calculations.  The reaction mechanism of the $f_1(1285)$
photoproduction considered in the current work is illustrated in Fig. 1.
\begin{figure}[tbph]
\begin{center}
\includegraphics[scale=0.54]{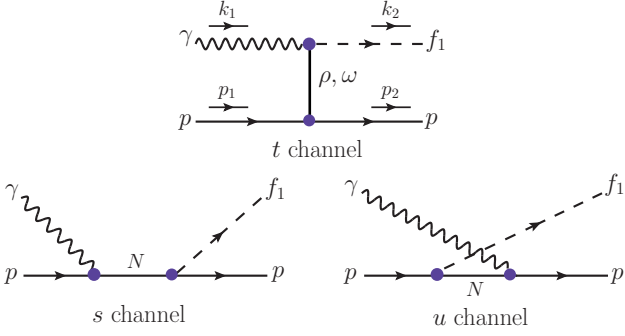}
\end{center}
\caption{Feynman diagrams for the $f_{1}(1285)$
photoproduction.}
\end{figure}

In Ref.~\cite{Kochelev:2009xz}, the $t$-channel $\rho $ and $\omega $ meson
exchanges were considered with the Reggeized trajectory, which failed to
reproduce the rapid falloff at forward angles in the CLAS data. In this
work, we will adopt  an interpolating Reggeized trajectory instead of the
traditional Reggeized treatment to reproduce the behavior of the experimental data
at forward angles~\cite{Nam:2010au}. Usually, the contribution from the $s$
channel with nucleon pole is expected to be very small and the $u$-channel
contribution will produce an enhancement at backward angles~\cite%
{He:2013ksa,He:2014gga}. we hope to reproduce the differential cross
section of the new CLAS data only with the $t$- and $u$-channel
contributions. Hence, in the present work, we do not include the
contributions from the nucleon resonances in the $s$ channel, which was
suggested in Ref.~\cite{Dickson:2016gwc} . A calculation with the major background
mechanisms may be enough to capture the main features of the data. Of
course, such assumptions need explicit calculation to confirm them, which will be
done in this work.

This paper is organized as follows. After the introduction, we present the
formalism including Lagrangians and amplitudes of the $f_1(1285)$
photoproduction in Sec. II. The numerical results of differential cross
section follow in Sec. III and are compared with the CLAS data. Finally, the
paper{\ ends} with a brief summary.

\section{Formalism}

\subsection{Lagrangians and amplitudes}

The $f_{1}(1285)$ photoproduction off a proton target occurs through the
mechanism in Fig. 1, which includes $t$-channel $\rho $ and $\omega $
exchanges, and $s$ and $u$ channels with intermediate nucleons. To gauge the
contributions from these mechanisms, the relevant Lagrangians are needed,
which will be given in the following.

For the $t$-channel vector-meson ($V=\rho $ or $\omega $) exchange, one needs the 
following Lagrangians \cite{Tsushima:1996xc,Tsushima:1998jz,Sibirtsev:2003ng,Kochelev:2009xz,Domokos:2009cq},
\begin{eqnarray}
\mathcal{L}_{VNN} &=&-g_{VNN}\bar{N}\left[ \gamma _{\mu }
{V}^{\mu }-\frac{\kappa _{V}}{2m_{N}}\sigma _{\mu \nu }\partial ^{\nu
}{V}^{\mu }\right]{N}, \\
\mathcal{L}_{Vf_{1}\gamma } &=&g_{Vf_{1}\gamma }\epsilon _{\mu \nu
\alpha \beta }\partial ^{\mu }A^{\alpha }\partial^2{V}^{\nu }f_{1}^{\beta },
\end{eqnarray}%
where $N$, $V$, $f_{1}$, and $A$ are the nucleon, vector meson, $f_{1}(1285)$
meson, and photon fields, respectively. The coupling constant $g_{\rho
f_{1}\gamma }$ is determined from the decay width
\begin{equation}
\Gamma _{f_{1}\rightarrow \rho \gamma }=g_{\rho f_{1}\gamma }^{2}\frac{%
m_{\rho }^{2}(m_{f_{1}}^{2}+m_{\rho }^{2})(m_{f_{1}}^{2}-m_{\rho }^{2})^{3}}{%
96\pi m_{f_{1}}^{5}}.
\end{equation}%
In the new CLAS experiment, the radiative decay of the $f_{1}(1285)$ to $\rho $
meson was extracted from the $f_1(1285)$ decay process, a decay width $\Gamma
_{f_{1}\rightarrow \rho \gamma }\simeq 453\pm 177$ keV was obtained, which
is much smaller than the PDG value of $1331\pm 320$ keV~\cite{Dickson:2016gwc}.
In the current work, the CLAS
value will be adopted in the calculation for
consistency, and one gets $g_{\rho f_{1}\gamma }\simeq 0.56$ GeV$^{-2}$. The
value of $g_{\omega f_{1}\gamma }$ can be obtained via quark model and SU(2)
symmetry \cite{Kochelev:2009xz}, i.e.%
\begin{equation*}
g_{\omega f_{1}\gamma }=\frac{1}{3}g_{\rho f_{1}\gamma }\text{.}
\end{equation*}%
In the literature \cite{Kochelev:2009xz,Sibirtsev:2003ng,Domokos:2009cq}, the correspondent coupling
constants $g_{VNN}$ and $\kappa _{V}$ have been calculated. In Table \ref%
{tab:vector}, we list the values of the coupling constants associated with $\rho $ and $%
\omega $ mesons which will be adopted in the calculation. \renewcommand\tabcolsep{0.8cm} \renewcommand{%
\arraystretch}{1.2}
\begin{table}[bph]
\caption{The values of coupling constants related to the vector mesons~\protect\cite%
{Kochelev:2009xz,Sibirtsev:2003ng,Domokos:2009cq}. }
\label{tab:vector}%
\begin{tabular}{|c|c|c|c|}
\hline\hline
$V$ & $g_{VNN}$ & $\kappa _{V}$ & $g_{Vf_{1}\gamma }$ \\ \hline
$\rho $ & 2.4 & 6.1 & 0.56 \\
$\omega $ & 9 & 0 & 0.18 \\ \hline
\end{tabular}%
\end{table}

According to the above Lagrangians, the scattering amplitude of the $%
f_{1}(1285)$ photoproduction via vector meson exchange can be written as%
\begin{eqnarray}
i\mathcal{M}_{t} &=&ig_{VNN}g_{Vf_{1}\gamma }F_{t}(q_{V})\epsilon
_{f_{1}}^{\nu \ast }(k_{2})\bar{u}(p_{2})q_{V}^{2}\epsilon _{\mu \nu \alpha
\beta }\frac{\mathcal{P}^{\alpha \xi }}{t-m_{V}^{2}}  \notag \\
&&\left( \gamma _{\xi }-i\frac{\kappa _{V}}{2m_{N}}\gamma _{\xi }%
\rlap{$\slash$}q_{V}\right) k_{1}^{\beta }u(p_{1})\epsilon ^{\mu }(k_{1}),
\label{AmpT}
\end{eqnarray}%
with propagator of the vector meson as $\mathcal{P}^{\alpha \xi }=i\left(
g^{\alpha \xi }+q_{V}^{\alpha }q_{V}^{\xi }/m_{V}^{2}\right) $. For the $t$-channel vector-meson exchange, the general form factors $F_{Vf_{1}\gamma
}=[(\Lambda _{t}^{2}-m_{V}^{2})/(\Lambda _{t}^{2}-q_{V}^{2})]^{2}$ and $%
F_{VNN}=(\Lambda _{t}^{2}-m_{V}^{2})/(\Lambda _{t}^{2}-q_{V}^{2})$ are taken
into account and in this work the cutoffs are taken as the same ones for
the simplification as done in Ref.~\cite{Kochelev:2009xz}. Here, $q_{V}$ and $m_{V}$
are the four-momentum and mass of the exchanged meson, respectively.

To study the contribution from the nucleon exchange, one needs to construct
the Lagrangians for the $\gamma NN$ and the $f_{1}NN$ couplings~\cite%
{Domokos:2009cq,Oh:2007jd},%
\begin{eqnarray}
\mathcal{L}_{\gamma NN} &=&\frac{e\kappa _{p}}{2m_{N}}\bar{N}\sigma
_{\mu \nu }\partial ^{\nu }A^{\mu }{N}+\text{H.c.}~, \\
\mathcal{L}_{f_{1}NN} &=&g_{f_{1}NN}\bar{N}\left( f_{1}^{\mu }-i\frac{\kappa _{f_{1}}}{2m_{N}}\gamma ^{\nu }\partial _{\nu
}f_{1}^{\mu }\right) \gamma _{\mu }\gamma ^{5}{N}+\text{%
H.c.}.
\end{eqnarray}%
The value of coupling constant $g_{f_{1}NN}$ is not well determined, and a
value of about 2.5 will be taken as discussed in Ref. ~\cite{Birkel:1995ct}. $\kappa
_{p}\simeq 1.79$ is the anomalous magnetic moment of the proton. The value
of $\kappa _{f_{1}}$ will be determined by fitting the CLAS experimental
data \cite{Dickson:2016gwc}. For the $s$ and $u$ channels with intermediate nucleons,
we adopt the{\ general form factor to describe the size of the hadrons \cite%
{Kochelev:2009xz},}%
\begin{equation}
F_{s/u}(q_{N})=\frac{\Lambda _{s/u}^{4}}{\Lambda
_{s/u}^{4}+(q_{N}^{2}-m_{N}^{2})^{2}}~,
\end{equation}%
where $q_{N}$ and $m_{N}$ are the four-momentum and mass of the exchanged
nucleon, respectively. Since the $s$-channel contribution is usually very
small, we take $\Lambda _{s}=\Lambda _{u}$. The values of cutoffs
$\Lambda _{u}$ and $\Lambda _{t}$ will be{\ determined by fitting
experimental data}.

The scattering amplitudes via the nucleon exchanges read as%
\begin{eqnarray}
i\mathcal{M}_{s} &=&\frac{e\kappa _{p}}{2m_{N}}g_{f_{1}NN}F_{s}(q_{N})%
\epsilon _{f_{1}}^{\nu \ast }(k_{2})\bar{u}(p_{2})\left( 1-i\frac{\kappa
_{f_{1}}}{2m_{N}}\rlap{$\slash$}k_{2}\right) \gamma _{\nu }\gamma ^{5}
\notag \\
&&\frac{(\rlap{$\slash$}q_{N}+m_{N})}{s-m_{N}^{2}}\gamma _{\mu }%
\rlap{$\slash$}k_{1}u(p_{1})\epsilon ^{\mu }(k_{1}), \\
i\mathcal{M}_{u} &=&\frac{e\kappa _{p}}{2m_{N}}g_{f_{1}NN}F_{u}(q_{N})%
\epsilon _{f_{1}}^{\nu \ast }(k_{2})\bar{u}(p_{2})\gamma _{\mu }%
\rlap{$\slash$}k_{1}\left( 1-i\frac{\kappa _{f_{1}}}{2m_{N}}\rlap{$\slash$}%
k_{2}\right)  \notag \\
&&\gamma _{\nu }\gamma ^{5}\frac{(\rlap{$\slash$}q_{N}+m_{N})}{u-m_{N}^{2}}%
u(p_{1})\epsilon ^{\mu }(k_{1}),
\end{eqnarray}%
where $s=(k_{1}+p_{1})^{2}$, $t=(k_{1}-k_{2})^{2}$, and $u=(p_{2}-k_{1})^{2}$
are the Mandelstam variables.

\subsection{Interpolating Reggeized $t$ channel}

To analyze hadron photoproduction at high energies, a more economical
approach may be furnished by a Reggeized treatment \cite{Haberzettl:2015exa}. In Refs.
\cite{Haberzettl:2015exa,Wang:2015hfm}, standard Reggeized treatment for $t$-channel meson
exchange consists of replacing the product of the form factor in Eq.~(\ref%
{AmpT}) with
\begin{eqnarray}
F_{t}(q_{\rho }) &\rightarrow &\mathcal{F}_{t}(q_{\rho })=(\frac{s}{s_{scale}%
})^{\alpha _{\rho }(t)-1}\frac{\pi \alpha _{\rho }^{\prime }(t-m_{\rho }^{2})%
}{\Gamma \lbrack \alpha _{\rho }(t)]\sin [\pi \alpha _{\rho }(t)]}, \\
F_{t}(q_{\omega }) &\rightarrow &\mathcal{F}_{t}(q_{\omega })=(\frac{s}{%
s_{scale}})^{\alpha _{\omega }(t)-1}\frac{\pi \alpha _{\omega }^{\prime
}(t-m_{\omega }^{2})}{\Gamma \lbrack \alpha _{\omega }(t)]\sin [\pi \alpha
_{\omega }(t)]}.
\end{eqnarray}%
The scale factor $s_{scale}$ is fixed at 1 GeV. In addition, the Regge
trajectories $\alpha _{\rho }(t)$ and $\alpha _{\omega }(t)$ read as \cite%
{Kochelev:2009xz,Laget:2005be},%
\begin{equation}
\alpha _{\rho }(t)=0.55+0.8t,\ \alpha _{\omega }(t)=0.44+0.9t.\quad \ \
\end{equation}

In practical applications, the onset of the \textquotedblleft Regge regime"
is often very much under debate. In this work, we adopt the
interpolating Reggiezed treatment which can interpolate the Regge case
smoothly to the Feynman case. Such a hybrid approach has been successfully
applied to reproduce the experimental data in Refs.~\cite%
{Nam:2010au,He:2012ud,He:2013ksa,He:2014gga}, especially the falloff at
forward angles. In Refs. \cite{Haberzettl:2015exa,Wang:2015hfm}, the local gauge invariance
of the Regge trajectory was discussed. In the case of this work, the gauge
invariance is kept because the Lagrangians for the $t$ channel produce an
amplitude which keeps gauge invariance by itself. So we need not make extra
treatment to restore the gauge invariance. The interpolated Reggeized form
factor can then be written as%
\begin{equation}
F_{t}\rightarrow \mathcal{F}_{R,t}=\mathcal{F}_{t}R\left( t\right) +F_{t}%
\left[ 1-R\left( t\right) \right]\label{Eq: intRe}
\end{equation}%
where $R\left( t\right) =$ $R_{s}\left( s\right) R_{t}\left( t\right) $, with%
\begin{equation}
R_{s}\left( s\right) =\frac{1}{1+e^{-(s-s_{R})/s_{0}}},\ \ R_{t}\left(
t\right) =\frac{1}{1+e^{-(t+t_{R})/t_{0}}}.
\end{equation}%
Here, $s_{R}$ and $t_{R}$ describe the centroid values for the transition
from non-Regge to Regge regimes while $s_{0}$ and $t_{0}$ provide the
respective widths of the transition regions. The four parameters of this
function will be fitted to the experimental data.

\section{Numerical results}

With the preparation in the previous section, the differential cross section
of the $f_1(1285)$ photoproduction will be calculated and compared with the
CLAS data released recently. The differential cross section in the center of
mass (c.m.) frame is written with the amplitudes obtained in the previous
section as
\begin{equation}
\frac{d\sigma }{d\cos \theta }=\frac{1}{32\pi s}\frac{\left\vert \vec{k}%
_{2}^{{~\mathrm{c.m.}}}\right\vert }{\left\vert \vec{k}_{1}^{{~\mathrm{c.m.}}%
}\right\vert }\left( \frac{1}{4}\sum\limits_{\lambda }\left\vert \mathcal{M}%
\right\vert ^{2}\right) ,
\end{equation}%
where $s=(k_{1}+p_{1})^{2}$, and $\theta $ denotes the angle of the outgoing
$f_{1}(1285)$ meson relative to beam direction in the c.m. frame. $\vec{k}%
_{1}^{{~\mathrm{c.m.}}}$ and $\vec{k}_{2}^{{~\mathrm{c.m.}}}$ are the
three-momenta of the initial photon beam and final $f_{1}(1285)$, respectively.

\subsection{Fitting procedure}

The CLAS data~\cite{Dickson:2016gwc} for the $f_{1}(1285)$ photoproduction will be
fitted with the help of the \textsc{minuit} code in the \textsc{cernlib}. In
the current work, we minimize $\chi ^{2}$ per degree of freedom ($d.o.f.$)
for the differential cross sections $d\sigma /d\cos \theta $ of the CLAS
data by fitting seven parameters, which include four parameters for the
Regge trajectory ($s_{0}$, $t_{0}$, $s_{R}$, $t_{R}$), the anomalous magnetic
moment $\kappa _{f_{1}}$, and the cutoffs $\Lambda _{t}$ and $\Lambda _{u}$.
Here the cutoff $\Lambda _{s}$ for the $s$ channel is chosen to be the same as $%
\Lambda _{u}$ for the $u$ channel for simplification because the $s$-channel
contribution is usually small. The CLAS experimental data include 45 data
points at center-of-mass energy bin $W$= 2.35 2.45, 2.55, 2.65, and 2.75 GeV
at nine angle bins~\cite{Dickson:2016gwc}. In the fitting, both statistical and
systematic uncertainties are considered.

\renewcommand\tabcolsep{0.28cm} \renewcommand{\arraystretch}{1.2}
\begin{table}[h]
\caption{Fitted values of free parameters and corresponding reduced $\protect%
\chi ^{2}/d.o.f.$ value.}
\label{tab:fit}%
\begin{tabular}{|c|c|c|c|}
\hline\hline
$s_{0}$ (GeV$^{2}$) & $s_{R}$ (GeV$^{2}$) & $t_{0}$ (GeV$^{2}$) & $t_{R}$
(GeV$^{2}$) \\
$3.99\pm 0.23$ & $6.61\pm 0.59$ & $0.95\pm 0.07$ & $0.3\pm 0.08$ \\ \hline
$\kappa _{f_{1}}$ & $\Lambda _{t}$ (GeV) & $\Lambda _{s}=\Lambda _{u}$ (GeV)
& $\chi ^{2}/d.o.f.$ \\
$1.94\pm 0.39$ & $0.92\pm 0.04$ & $0.68\pm 0.06$ & $1.27$ \\ \hline
\end{tabular}%
\end{table}

The fitted values of the free parameters are listed in Table \ref{tab:fit},
with a reduced value $\chi ^{2}/d.o.f.=1.27$, which indicates the CLAS \cite%
{Dickson:2016gwc} data can be reproduced quite well in the current model with only $s$%
-, $u$-, and $t$-channel Born terms. It suggests that the intermediate $s$%
-channel nucleon resonances are not essential to reproduce the experimental
data. Moreover, the best fitted result is achieved with reasonable cutoff
values $\Lambda _{x}(x=s,u,t)$ around the usual empirical 1 GeV value.

We also make a fitting with the coupling constant of the radiative decay of
the $f_1(1285)$ $g_{\rho f_{1}\gamma }$ as a free parameter to test the
radiative decay width we adopted to calculate the coupling constant.
The fitted values of free parameters are listed in Table~\ref{Tabl:
parafree}. One
can obtain a similar result with a little smaller $\chi ^{2}/d.o.f.=1.21$.
The parameters are close to these with fixed $g_{\rho f_{1}\gamma }=0.56$ GeV%
$^{-2}$. The corresponding decay width $\Gamma _{f_{1}\rightarrow \rho
\gamma }\simeq 259\pm 12$ keV, which is within the margin of the CLAS
experimental value. Compared with the results with the coupling constant fixed at the
CLAS value, we can say that our fitting of the differential cross section is consistent with
the radiative decay of the $f_1(1285)$ observed by the CLAS Collaboration.

\renewcommand\tabcolsep{0.28cm} \renewcommand{\arraystretch}{1.2}
\begin{table}[h]
\caption{Fitted values of free parameters with the coupling constant
$g_{\rho f_1\gamma}$ as a free parameter.  Here, $\protect\chi ^{2}/d.o.f.=1.21
$.\label{Tabl: parafree}}
\label{tab:fitIII}%
\begin{tabular}{|c|c|c|c|}
\hline\hline
$s_{0}$ (GeV$^{2}$) & $s_{R}$ (GeV$^{2}$) & $t_{0}$ (GeV$^{2}$) & $t_{R}$
(GeV$^{2}$) \\
$3.99\pm 0.57$ & $5.00\pm 0.77$ & $0.92\pm 0.02$ & $0.3\pm 0.07$ \\ \hline
$\kappa _{f_{1}}$ & $\Lambda _{t}$ (GeV) & $\Lambda _{s}=\Lambda _{u}$ (GeV)
& $g_{\rho f_{1}\gamma }$ \\
$1.92\pm 0.06$ & $0.87\pm 0.02$ & $0.68\pm 0.01$ & $0.42\pm 0.01$ \\ \hline
\end{tabular}%
\end{table}

\subsection{\boldmath Differential cross section for the $f_{1}(1285)$
photoproduction}

\begin{figure}[h]
\centering
\includegraphics[scale=0.4]{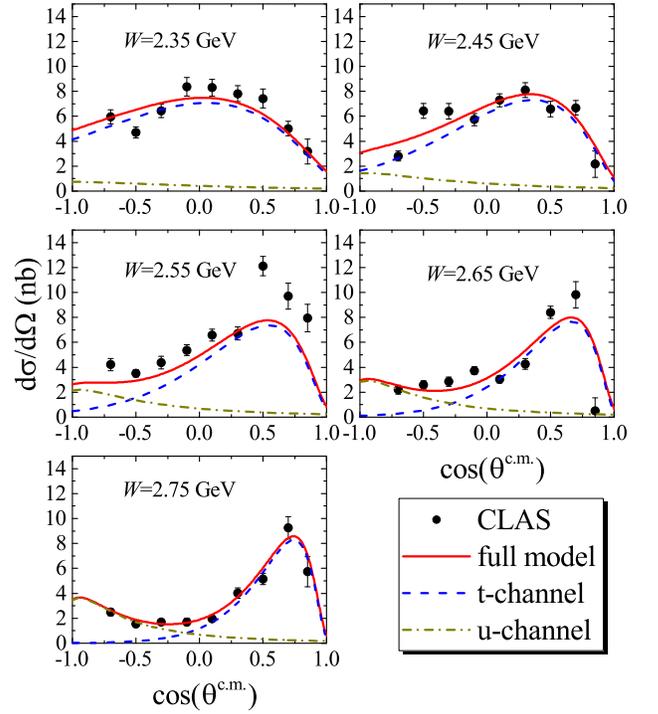}
\caption{The differential cross section $d\protect\sigma %
/d\cos \protect\theta $ for the $f_{1}(1285)$ photoproduction off a proton
as a function of $\cos \protect\theta $. The data are from Ref. \protect\cite%
{Dickson:2016gwc}. The full (red), dashed (blue), dash-dotted (dark yellow) and
dash-dot-dotted (green) lines are for the full model, $t$ channel, $u$ channel
and Feynman model, respectively. The curves have been scaled by the PDG
\protect\cite{Olive:2016xmw} branching fraction for $f_{1}(1285)\rightarrow
\protect\eta \protect\pi ^{+}\protect\pi ^{-}$. In the figures, only
statistical uncertainty is presented as in Ref.~\protect\cite{Dickson:2016gwc} }
\label{fig:data}
\end{figure}

In this subsection, we will present the best fitted results with the
fixed coupling constant $g_{f_1 \rho \gamma}$. The fitted values of the free parameters are listed in Table \ref{tab:fit},
and the reduced value $\chi ^{2}/d.o.f.=1.27$.  As shown in
Fig.~\ref{fig:data}, the differential cross section ${d\sigma }/{\ d\cos
\theta }$ reported by the CLAS Collaboration is well reproduced in our
model. It is found that the $t$-channel vector-meson exchange plays a
dominant role in the $f_1(1285)$ photoproduction. Its contribution
corresponds to the enhancement at forward angles. The $u$ channel with
nucleon exchange plays a very important role at backward angles, in
particular, at the high-energy end of the data range. The differential cross
section in the full model is almost from these two contributions. Since the
contribution from the $s$ channel with nucleon exchange is very small and can be
negligible, it is not shown in Fig. \ref{fig:data}.

To show the importance of introducing the interpolated Reggeized
form factor in Eq.~(\ref{Eq: intRe}), in Fig.~\ref{fig:data} we
present the results in the  Feynman model, in which the
Feynman $t$ channel without inclusion of the interpolated Reggeized form factor, $u$ channel and $s$ channel are considered.  The best fitted parameters are
listed in Table \ref{Tab: paraFeymann}. A very large $\chi^2/d.o.f$ of
6.31 suggests that the differential cross section observed at CLAS cannot
be reproduced from the  Feynman model.  The most serious defect is that the
sharp falloff at forward angle cannot be reproduced, which leads to
the difficulty of reproducing the magnitude of the differential cross
section. At the backward angles, the $u$-channel contribution shows its
importance to reproducing the experimental data. After the interpolated Reggeized form factor is introduced, the CLAS
data are well reproduced, as shown in  Fig. \ref{fig:data}.
\renewcommand\tabcolsep{0.28cm} \renewcommand{\arraystretch}{1.2}
\begin{table}[h]
\caption{Fitted values of free parameters with the Feynman model.
Here $\protect\chi ^{2}/d.o.f.=6.31
$.\label{Tab: paraFeymann}}%
\begin{tabular}{|c|c|c|c|}
\hline\hline
$\kappa _{f_{1}}$ & $\Lambda _{t}$ (GeV) & $\Lambda _{s}=\Lambda _{u}$ (GeV)
& $g_{\rho f_{1}\gamma }$ \\
$0.95\pm 0.52$ & $0.98\pm 0.01$ & $0.50\pm 0.02$ & $0.99\pm 0.01$ \\ \hline
\end{tabular}%
\end{table}

In the work of Kochelev $et\ al.$, the Reggeized treatment has been considered in the $t$ channel.
Because there was no experimental data when Kochelev $et\ al.$ made the
prediction about this process, the current results are much closer to the
experimental data than the prediction in Refs.~\cite{Kochelev:2009xz,Dickson:2016gwc}. The
main improvements in our model are the inclusion of the interpolating Reggeized
treatment and the inclusion of the $u$-channel contribution. The $u$-channel
contribution is essential to reproducing the enhancement at backward angles,
which is absent in Refs.~\cite{Kochelev:2009xz,Dickson:2016gwc}.   Kochelev $et\ al.$ adopted  an original Reggeized treatment without interpolation, so the
differential cross section was predicted to be smaller than the experimental data at low energies, while 
the cross section at high energy was predicted closely~\cite%
{Kochelev:2009xz,Dickson:2016gwc}. The interpolating Reggeized treatment makes it possible
to produce the relative magnitude of the differential cross section at both
low and high energies. Further, the rapid falloff at forward angles can also be
reproduced with the interpolating Reggiezed treatment.

\begin{figure}[h]
\begin{center}
\includegraphics[scale=0.4]{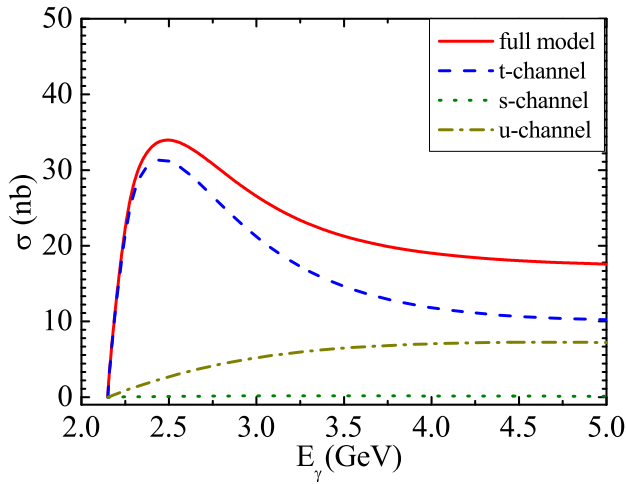}
\end{center}
\caption{Total cross section for $\protect\gamma p\rightarrow
f_{1}p$ reaction. The full (red), dashed (blue), dotted (green) and
dash-dotted (dark yellow) lines are for the full model, $t$ channel, $s$
channel and $u$ channel, respectively.}
\label{Fig:total}
\end{figure}

We also present the total cross section of the $f_{1}(1285)$ photoproduction
in Fig.~\ref{Fig:total}. It is found that the $t$-channel vector-meson
exchange is dominant at beam energy $E_{\gamma }$ from threshold up to 5.0
GeV. The contribution from the $u$ channel increases with the increase of
energy. The contribution from the $s$-channel nucleon exchange is small and can
be neglected. The total cross section of the $f_{1}(1285)$ photoproduction
could be significant at higher energies.

\section{Summary and discussion}

Within an effective Lagrangian approach, a seven-parameter fitting is done
to the CLAS data of the $f_{1}(1285)$ photoproduction off a proton with an
interpolating Reggeized treatment. The numerical results show that the
differential cross section is well reproduced by our model with $\chi
^{2}/d.o.f.=1.27$. The fitting value of the decay width $\Gamma
_{f_{1}\rightarrow \rho \gamma }$ with the differential cross section is
also consistent with the CLAS value.

The numerical results suggest that the $t$ channel is dominant in the reaction
mechanism of the $f_1(1285)$ photoproduction and responsible for the behavior
of the cross sections at forward angles. As expected, the contribution from the $%
s$-channel nucleon exchange is so small that it almost has no effect on the
differential cross sections. The CLAS \cite{Dickson:2016gwc} data cannot be fitted
very well using the usual Feynman-type $t$-channel exchange alone, even with
the traditional Reggeized treatment, which suggests that the interpolating
Reggeized treatment of the $t$ channel is essential to achieve the fit
quality exhibited in Fig. \ref{fig:data}. Further, the $u$ channel is
responsible for the enhancement of the differential cross section at backward
angles.

In the current work, $s$-channel intermediate nucleon resonances are not
included in the fitting of the experimental data. The small $\chi ^{2}$
suggests their contribution should not be very large, which indicates weak
couplings of the $f_{1}(1285)$ and the nucleon resonances in the energy
region considered in this work. Though the results of the differential cross
section supported the small contribution from the nucleon resonance, it
should be explained why there is a nucleon-resonance signal in the
Dalitz plot. A explicit calculation is out of the scope of the current work.
However, we would like to give some arguments to explain why an $s$-channel
effect can be found within the $t$-channel-dominant interaction mechanism. In
interpolating Reggeized treatment, we need  to introduce an auxiliary function
to interpolate the Regge case and Feynman case as shown in Eqs. (13) and (14).
The $R_{s}(s)$ will introduce an $s$ dependence of the cross section. If we
expand the exponential function (even with other forms) at $s_{R}$, we have $%
R_{s}(s)=1/(1+1-(s-s_{R})/s_{0})=-s_{0}/(s-s_{R}-2s_{0})$. Obviously, at the
energies around  $\sqrt{s_{R}}$, the amplitudes will exhibit some
characteristics of the $s$ channel.

To describe more detailed structures of the cross sections of $f_{1}(1285)$
photoproduction, inclusion of $s$-channel resonances may be necessary, For
example, some data points at forward angles with $W=2.55$ GeV are larger
than the theoretical result. However, the current data is not enough to make a
meaningful analysis. To warrant expanding efforts in this direction, more
precise data for $f_{1}(1285)$ photoproduction are necessary. The near
future CLAS12@JLab experiment \cite{Fegan:2014dba,Glazier:2015cpa} may provide great
opportunities for research in this direction.

\section{Acknowledgments}

This project is supported by the National Natural Science Foundation of
China under Grant No. 11675228 and the Major State Basic Research Development
Program in China under Grant no. 2014CB845405.

\end{document}